\documentclass{Interspeech2024}

\interspeechcameraready

\title{FLY-TTS: Fast, Lightweight and High-Quality End-to-End Text-to-Speech Synthesis}

\name[affiliation={}]{Yinlin}{Guo$^\heartsuit$}
\name[affiliation={}]{Yening}{Lv$^\heartsuit$}
\name[affiliation={}]{Jinqiao}{Dou}
\name[affiliation={}]{Yan}{Zhang}
\name[affiliation={}]{Yuehai}{Wang$^\diamondsuit$}

\address{
  College of Information Science and Electronic Engineering, Zhejiang University, China
}
\email{\{22231138, 22231057, 3170105596, 22231136, wyuehai\}@zju.edu.cn}


\usepackage{multirow}
\usepackage{hyperref}
\hypersetup{hidelinks,
	colorlinks=true,
	allcolors=black,
	pdfstartview=Fit,
	breaklinks=true}
\begin{document}

\maketitle
\renewcommand{\thefootnote}{\fnsymbol{footnote}}
\footnotetext{$^\heartsuit$Equal contribution}
\footnotetext{$^\diamondsuit$Corresponding author}
\renewcommand{\thefootnote}{\arabic{footnote}}
\begin{abstract}

    While recent advances in Text-To-Speech synthesis have yielded remarkable improvements in generating high-quality speech, research on lightweight and fast models is limited.
    This paper introduces FLY-TTS, a new fast, lightweight and high-quality speech synthesis system based on VITS.
    Specifically, 1) We replace the decoder with ConvNeXt blocks that generate Fourier spectral coefficients followed by the inverse short-time Fourier transform to synthesize waveforms;
    2) To compress the model size, we introduce the grouped parameter-sharing mechanism to the text encoder and flow-based model;
    3) We further employ the large pre-trained WavLM model for adversarial training to improve synthesis quality.
    Experimental results show that our model achieves a real-time factor of 0.0139 on an Intel Core i9 CPU, 8.8x faster than the baseline (0.1221), with a 1.6x parameter compression.
    Objective and subjective evaluations indicate that FLY-TTS exhibits comparable speech quality to the strong baseline.\footnote{Audio samples are available at \href{https://lilyn3125.github.io/flytts}{https://lilyn3125.github.io/flytts}}
\end{abstract}

\section{Introduction}
\vspace{-0.05in}
Text-to-speech (TTS) synthesis is the process of converting input text into speech.
Recent advances in TTS synthesis have achieved significant improvements in high-quality speech synthesis \cite{app9194050,tan2021survey}.
Most of the work has focused on improving the quality and naturalness of synthesized speech \cite{tacotron,glow-tts,vits,naturalspeech}.
However, to apply TTS models in real-world applications typically faces three challenges:
1) Current TTS models are usually too large to deploy on edge or mobile devices;
2) The slow inference of the model limits its application in low computing resources;
3) In general, larger model sizes tend to improve performance, and there is often a trade-off between the two factors.

To address these challenges, in this paper, we propose \textbf{FLY-TTS}, a new \textbf{F}ast, \textbf{L}ightweight and high-qualit\textbf{Y} \textbf{T}ext-\textbf{T}o-\textbf{S}peech synthesis system.
Based on VITS \cite{vits}, FLY-TTS demonstrates efficient inference capabilities and model size reduction, while maintaining comparable synthesis quality.
Recent research has shown that the HiFi-GAN-based decoder is the main bottleneck of VITS with respect to inference speed \cite{mb-istft-vits}.
Therefore, we suggest to use ConvNeXt blocks to speed up inference by generating Fourier spectral coefficients and reconstructing raw waveforms directly through the inverse short-time Fourier transform.
To reduce the model size, we introduce the grouped parameter-sharing mechanism into the text encoder and flow-based model in VITS.
By adjusting the number of groups, we can balance between model size and modeling capacity.
To mitigate the decline in synthesis quality caused by model compression, we introduce a large pre-trained WavLM \cite{Wavlm} model as a discriminator for adversarial training.
By leveraging self-supervised representation to furnish the generator with additional information, we aim to enhance the quality of synthesized speech.
Experiments on the LJSpeech dataset \cite{ljspeech17} show that FLYTTS achieves a real-time factor of 0.0139 on Intel Core i9 CPU, 8.8 times faster than the baseline system (0.1221), and the parameters are compressed by 1.6 times.

\section{Related work}

\subsection{End-to-end TTS}
End-to-end TTS directly synthesizes speech waveforms from text without explicit intermediate representations.
It has gradually shifted from autoregressive models to non-autoregressive models.

Autoregressive TTS models \cite{shen2018natural, tacotron} generate the next time-step output conditioned on previous frames, leading to large inference latency and robustness issues.
Recent researches have focused on non-autoregressive TTS models \cite{fastspeech, glow-tts, vits, naturalspeech},
which generate mel-spectrograms or raw waveforms in parallel.
FastSpeech \cite{fastspeech} and FastSpeech2 \cite{fastspeech2} parallelize the generation of mel-spectrograms by predicting the length of the mel-spectrograms with external alignment models or tools.
Glow-TTS \cite{glow-tts} combines flow models and dynamic programming to search for the most likely monotonic alignment between text and speech latent representations to achieve parallel TTS.
To further enhance the quality of synthesized speech, VITS \cite{vits} adopts variational inference with normalizing flows to generate raw waveforms.
The recently proposed NaturalSpeech \cite{naturalspeech} also uses a variational autoencoder (VAE) for end-to-end text-to-waveform generation, achieving human-level speech quality.
While these models have made notable advancements in synthesis quality, they typically suffer from inefficient inference, impeding their deployment in real-world scenarios.

\subsection{Lightweight and fast TTS}
Lightweight TTS aims to reduce model size and computational complexity while maintaining the quality of synthesized speech as much as possible.
While some recent works \cite{NixTTS, Light-tts, SpeedySpeech, Lightspeech, EfficientSpeech, AdaVITS, mb-istft-vits} have shown progress in this direction, they could encounter certain challenges.
Nix-TTS \cite{NixTTS}, Light-TTS \cite{Light-tts}, and SpeedySpeech \cite{SpeedySpeech} employ knowledge distillation to reduce parameters while hindering end-to-end training.
LightSpeech \cite{Lightspeech} resorts to neural architecture search (NAS) to design lightweight models but requires huge computational resources during training.
EfficientSpeech \cite{EfficientSpeech} adopts a lightweight U-Network to reduce parameters, but the synthesis quality still has room for improvement.
The most similar work to ours may be AdaVITS \cite{AdaVITS}, which proposes some tricks to achieve lightweight TTS.
However, it is tailored for speaker adaptation and incorporates PPG as a linguistic feature, introducing an additional PPG predictor module.
MB-iSTFT-VITS \cite{mb-istft-vits} also uses iSTFT to synthesize raw waveforms, but the decoder still contains computationally intensive upsampling operations.
Our proposed FLY-TTS exploits the large speech model as a discriminator to enjoy the benefits of self-supervised representation without affecting the generator.

\section{FLY-TTS}

\begin{figure*}[hbtp]
    \centering
    \includegraphics[width=\linewidth]{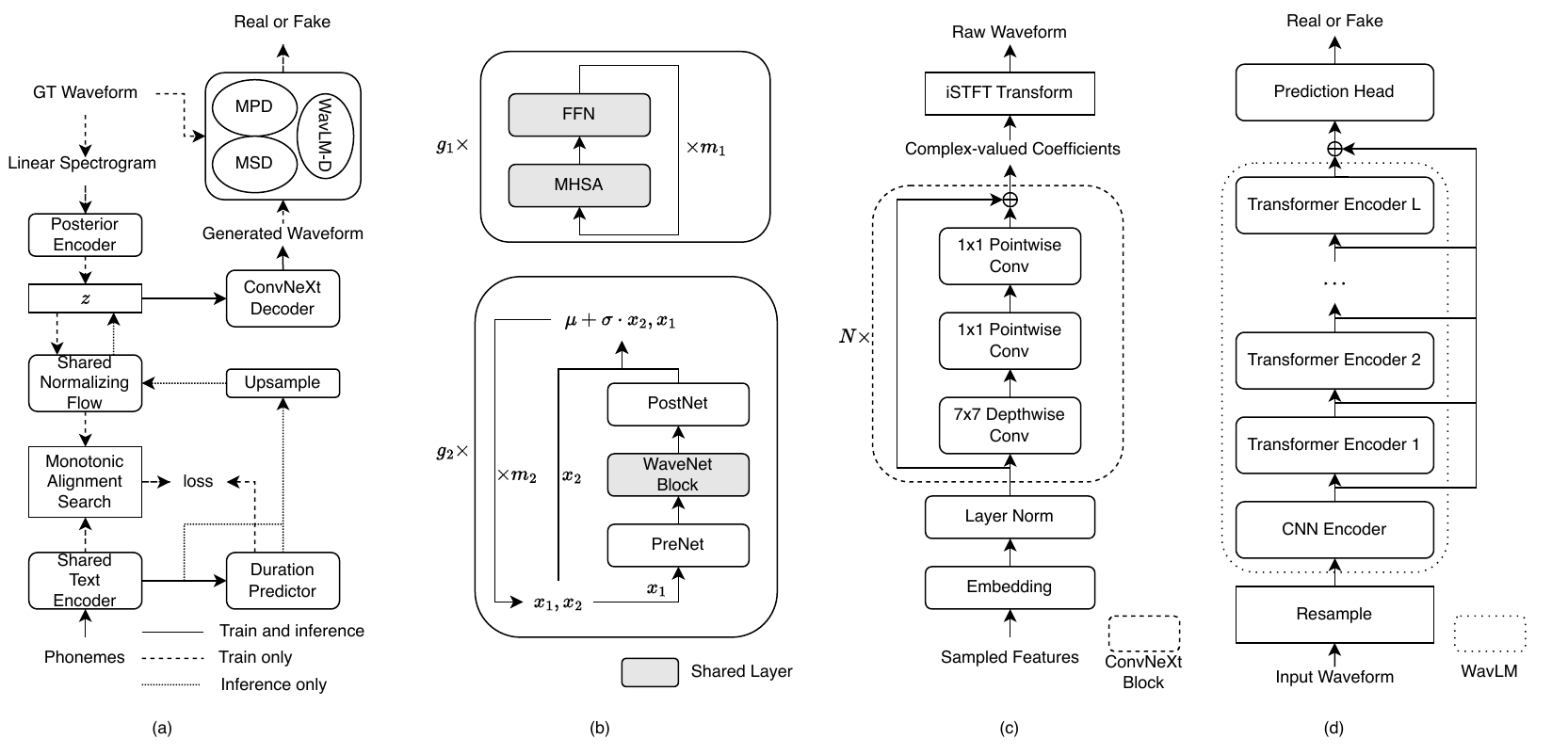}
    \caption{
        Our proposed FLY-TTS.
        (a): The overall architecture of FLY-TTS.
        (b): Text encoder (top half) and flow-based model (bottom half) with grouped parameter-sharing.
        (c): ConvNeXt-based decoder.
        (d): Large pre-trained WavLM model for adversarial training.
    }
    \label{fig:flytts}
\end{figure*}

As illustrated in Figure~\ref{fig:flytts}, our proposed FLY-TTS is based on the end-to-end VITS \cite{vits}, and the following briefly introduces the VITS.

VITS is a conditional VAE with the objective of maximizing the variational lower bound of the log-likelihood $\log p_\theta(x|c)$ of the target data $x$ given the input condition $c$:
$$\begin{aligned}
        \log p_\theta(x|c)\geq\mathbb{E}_{q_\phi(z|x)}\bigg[\log p_\theta(x|z){-}\log\frac{q_\phi(z|x)}{p_\theta(z|c)}\bigg],
    \end{aligned}$$
where $z$ is the latent variable, $p_\theta(z|c)$ is the prior distribution of $z$ given condition $c$, $p_\theta(x|z)$ is the likelihood of the data given $z$, and $q_\phi(z|x)$ is the approximate posterior distribution.
The model comprises a posterior encoder, prior encoder, and decoder,
corresponding to $q_\phi(z|x)$, $p_\theta(z|c)$, and $p_\theta(x|z)$, respectively.
Additionally, sets of discriminators are employed for adversarial training to enhance the quality of the synthesized speech.

\textbf{Prior encoder}: The prior encoder $E_{{\text{prior}}}$ receives input phonemes $c$ and predicts the prior distribution.
It consists of a text encoder for input processing and a normalizing flow $f_\theta$ that improves the flexibility of the prior distribution.

\textbf{Posterior encoder}: The posterior encoder $E_{{\text{posterior}}}$ operates on the linear spectrum to extract the mean and variance of the approximate posterior distribution.
Since this module is only used for training and will not impact the inference speed, no modifications are made to the posterior encoder in our proposed methods.

\textbf{Decoder}: The decoder $E_{{\text{decoder}}}$ generates waveforms from the latent $z$. This module is essentially the HiFi-GAN \cite{Hifi-gan} generator.

\textbf{Discriminators}: The discriminators $D$ in VITS are similar to those in HiFi-GAN \cite{Hifi-gan}.
They includes a set of multi-period discriminators and multi-scale discriminators, to improve the quality of the synthesized audio via adversarial training.

The proposed FLY-TTS shares similarities with VITS, as depicted in Figure~\ref{fig:flytts}a.
We outline the improvements of FLY-TTS in the following sections.

\subsection{Grouped parameter-sharing}

Parameter sharing is a widely used technique to improve parameter efficiency.
To achieve a trade-off between model size and expressiveness power,
we introduce the grouped parameter-sharing mechanism to the text encoder and flow-based model in the prior encoder.

The original text encoder in VITS is a multi-layer transformer encoder \cite{transformer}.
Previous work has shown that there exists redundancy in the transformer layers, and parameter-sharing could effectively reduce model size without substantially compromising performance \cite{ALBERT,universal}.
As shown in the upper half of Figure~\ref{fig:flytts}b, we adopt a sequential grouped parameter-sharing strategy,
assigning the same parameters to the sequential $m_1$ layers, with a total of $g_1 \times m_1$ layers, where $g_1$ is the number of groups.
When $g_1=1$, the grouped parameter-sharing mechanism becomes complete parameter sharing.

Flow-based models also suffer from the problem of large memory footprints.
Similar to the grouped parameter-sharing in the text encoder, we divide the $K=g_2\times m_2$ steps of flow $\mathbf{f}_1,\mathbf{f}_2, \cdots, \mathbf{f}_K$ into $g_2$ groups, each group containing $m_2$ flow steps.
Inspired by \cite{Portaspeech, AdaVITS},
we do not share the parameters of all modules in affine coupling layers.
Instead, following NanoFlow \cite{Nanoflow}, we only share the parameters of the projection layer composed of WaveNet \cite{WaveNet} (as shown in the lower half of Figure~\ref{fig:flytts}b),
while maintaining parameter independence across other modules.

\subsection{ConvNeXt based decoder}

The decoder in VITS is based on the HiFi-GAN vocoder,
which relies on transposed convolution to synthesize raw waveforms from the representation $z$.
Due to the time-consuming nature of the upsampling process, it is the main bottleneck in terms of inference speed \cite{mb-istft-vits}.

To this end, as shown in Figure~\ref{fig:flytts}c,
we draw inspiration from Vocos \cite{Vocos} and use ConvNeXt blocks \cite{convnext} as the foundational backbone to generate the Fourier time-frequency coefficients with the same temporal resolution.
Subsequently, we synthesize the raw waveforms through the fast inverse Short Time Fourier transform (iSTFT) to greatly reduce the computational cost.
The ConvNeXt module comprises a 7x7 depthwise convolution followed by a reverse bottleneck consisting of two 1x1 pointwise convolutions,
where the bottleneck employs GELU (Gaussian Error Linear Unit) activation.

Specifically, given the latent variable $z$, we first sample to obtain the feature sequence $S = [s_1, s_2, \cdots, s_T], s_i \in \mathbb{R}^{D}$,
where $D$ is the dimension of the hidden representation, and $T$ is the number of acoustic frames.

The features are then passed through the embedding layer to match the number of frequency bins ($N$) of the iSTFT.
A stacked layers of ConvNeXt blocks is then used to generate the Fourier time-frequency coefficients $M = [m_1, m_2, \cdots, m_T], P = [p_1, p_2, \cdots, p_T]$,
where $m_i \in \mathbb{R}^{N}$ is the amplitude of the complex Fourier coefficient, and $p_i \in \mathbb{R}^{N}$ is the phase:
$$[M,P] = \text{ConvNeXts}(\text{Embed}(S)).$$
The iSTFT transform is then used to get the generated waveforms $\hat{y}$:
$$\hat{y} = \text{iSTFT}(M, P).$$
In practice, the fast Fourier transform (FFT) algorithm is used to implement the iSTFT.
Since the temporal resolution $T$ of the Fourier transform coefficients is much smaller than the number of samples in raw waveforms,
there is a significant potential for accelerating the synthesis speed.

\subsection{Pre-trained speech model for adversarial training}
\label{sec:wavlm}
Pre-trained large speech models have been proven to contain rich acoustic and semantic information,
facilitating high-quality speech synthesis \cite{valle, Audiolm}.
However, applying large pre-trained speech models in the generator network typically involves substantial computational overhead, which is not suitable for fast synthesis.

We circumvent this problem by exploiting a pre-trained WavLM \cite{Wavlm} model as a discriminator for adversarial training.
This brings two benefits: 1) it can use the rich acoustic and semantic information learned by the self-supervised model to update the generator,
and 2) it can avoid impacting the model size and inference speed of the generator.

As shown in Figure~\ref{fig:flytts}d, the WavLM \cite{Wavlm} model is a speech self-supervised model that employs Wav2vec2 \cite{wav2vec2} as its backbone and consists of a convolutional feature encoder and Transformer encoders.
The speech waveforms are first resampled to 16kHz, followed by the extraction of intermediate features by WavLM.
The prediction head then performs discriminative prediction based on the features.
We follow VITS to use the least-squares loss function \cite{mao2017least} as the additional adversarial loss:
$$\begin{aligned}
         & L_{adv}(D_{\text{W}}) =\mathbb{E}_{(y,z)}\Big[(D_{\text{W}}(y)-1)^2+(D_{\text{W}}(\hat{y}))^2\Big], \\
         & L_{adv}(G) =\mathbb{E}_z\bigg[(D_{\text{W}}(\hat{y})-1)^2\bigg],
    \end{aligned}$$
where $D_{\text{W}}$ is the WavLM discriminator, $G$ is the generator of FLY-TTS, $y$ is the real speech, and $\hat{y}=G(z)$ is the synthesized speech.

The design of the prediction head is inspired by StyleTTS2 \cite{Styletts2}, comprising a series of convolutional networks with Leaky ReLU activation.
To mitigate the computational overhead brought by WavLM, we fix the parameters of WavLM and only update the prediction head, which also reduces the risk of overfitting.

\section{Experiments and results}

\subsection{Experimental setup}
We evaluate FLY-TTS on LJSpeech dataset \cite{ljspeech17}, which contains 13100 English audio clips and corresponding text transcripts.
The dataset has a total length of about 24 hours, and each audio file is a single-channel 16-bit PCM WAV with a sample rate of 22050 Hz.
Following the setup of \cite{mb-istft-vits}, the dataset is randomly divided into 12500 training samples, 100 validation samples, and 500 test samples.

\subsection{Model configuration}

We use the original VITS and MB-iSTFT-VITS \cite{mb-istft-vits} as our strong baseline models, denoted as VITS-base and MB-iSTFT-base, respectively.
In FLY-TTS, the ConvNext module is implemented based on Vocos\footnote{https://github.com/gemelo-ai/vocos}.
For WavLM, we use the WavLM-base model\footnote{https://huggingface.co/microsoft/wavlm-base-plus} pre-trained on 94k hours of speech data to initialize the parameters of the discriminator.
To study the performance of the proposed model with fewer parameters, we also trained a Mini FLY-TTS model by reducing the number of layers.
The configuration of each model is described as follows.

\textbf{VITS-base}: We use the official implementation\footnote{https://github.com/jaywalnut310/vits}.
The number of layers of the transformer in the text encoder is 6, and the number of flow steps is 4.

\textbf{MB-iSTFT-base}: Follow the official implementation of iSTFT-VITS\footnote{https://github.com/MasayaKawamura/MB-iSTFT-VITS} with multi-band configuration,
the number of sub-bands is 4, the \textit{nfft} off iSTFT is 16, and the \textit{hop length} is 4.
We keep other hyperparameters consistent with VITS-base.

\textbf{FLY-TTS}: As mentioned before, we set the hyperparameter of the grouped parameter-sharing in the text encoder to $g_1=2, m_1=3$, and the setting in the flow-based model is $g_2=2, m_2=2$.
The number of ConvNext modules in the decoder is 6, and the \textit{nfft} and \textit{hop length} of iSTFT transform are consistent with mel-spectrogram extraction settings, which are 1024 and 256, respectively.

\textbf{Mini FLY-TTS}: The small version of FLY-TTS, we set $g_1=g_2=1$ for full parameter sharing.
To match the capacity of the prior encoder, we reduce the number of ConvNext modules in the decoder to 4.

\subsection{Training and evaluation}
We use two NVIDIA RTX GeForce 3090 GPUs for training and the batch size is 64.
All models are trained for 800K steps.
Following the setup of \cite{mb-istft-vits}, we use the AdamW optimizer \cite{adamw} with $\beta_1=0.8, \beta_2=0.99$, weight decay is $0.01$, and the initial learning rate is set to $1 \times 10^{-4}$.
In each epoch, the learning rate decay is scheduled as a factor of $0.999^{1/8}$.

To evaluate the quality of the synthesized speech, a mean opinion score (MOS) test is conducted.
Raters are asked to listen to the audio samples from the test set and evaluate their naturalness using a 5-point scale ranging from 1 to 5.
Each audio sample is evaluated by at least 15 raters.
To measure the similarity between the synthesized speech and the ground truth, we use dynamic time warping to calculate the mel-cepstral distortion (MCD) weighted by speech length\footnote{https://github.com/chenqi008/pymcd}.
We also compute the word error rate (WER) with the Whisper medium \cite{Whisper} ASR system.
The model size and inference speed are evaluated using the number of parameters and the real-time factor (RTF).
We calculate the average RTF on Intel Core i9-10920X CPU @3.50GHz and NVIDIA GeForce RTX 3090 GPU, and no optimization techniques are used during the inference\footnote{Test conditions: Ubuntu 20.04, Python 3.9.18, PyTorch version: 2.1.0+cu121}.
All metrics are calculated on randomly selected samples.
\begin{table}[th]
    \caption{Comparison of model size and average RTF on Intel Core i9 CPU (RTF-CPU) and NVIDIA 3090 GPU (RTF-GPU).}
    \label{tab:obj_comparison}
    \centering
    \begin{tabular}{cccc}
        \toprule
        \multirow{2}{*}{\textbf{Model}} & \multirow{2}{*}{\#\textbf{Params}} & \multicolumn{2}{c}{\textbf{RTF}}            \\
        \cmidrule(lr{0pt}){3-4}
                                        &                                    & CPU                              & GPU      \\
        \midrule
        VITS-base                       & $28.11$ M                          & $0.1221$                         & $0.0276$ \\
        MB-iSTFT-base                   & $27.49$ M                          & $0.0274$                         & $0.0095$ \\
        \midrule
        FLY-TTS                         & $17.89$ M                          & $0.0139$                         & $0.0062$ \\
        Mini FLY-TTS                    & $10.92$ M                          & $0.0127$                         & $0.0061$ \\
        \bottomrule
    \end{tabular}
\end{table}
\subsection{Results}
\subsubsection{Model performance}

\begin{table}[th]
    \caption{Subjective and objective evaluation, including MCD, WER (in percentages), and MOS with 95\% confidence intervals (CI).}
    \label{table:mos_comparison}
    \centering
    \begin{tabular}{cccc}
        \toprule
        \textbf{Model} & \textbf{MCD} & \textbf{WER} & \textbf{MOS}(CI) \\
        \midrule
        Ground truth   & -            & $1.56$       & $4.21(\pm 0.10)$ \\
        VITS-base      & $5.49$       & $1.71$       & $4.15(\pm 0.09)$ \\
        MB-iSTFT-base  & $5.57$       & $1.89$       & $4.08(\pm 0.11)$ \\
        \midrule
        FLY-TTS        & $5.56$       & $1.77$       & $4.12(\pm 0.09)$ \\
        Mini FLY-TTS   & $5.63$       & $2.09$       & $4.05(\pm 0.09)$ \\
        \bottomrule
    \end{tabular}
\end{table}

Table \ref{tab:obj_comparison} shows the comparison of our proposed models with the strong baselines.
As results show, our proposed FLY-TTS and Mini FLY-TTS demonstrate significant reductions in model size.
Compared to VITS-base, FLY-TTS reduces parameters by approximately 36.4\%, while Mini FLY-TTS achieves an even smaller model size, with a reduction of 61.2\%.

The results also highlight the efficiency of our proposed models.
Specifically, FLY-TTS achieves an impressive RTF of 0.0139 on CPU and 0.0062 on GPU, with a speedup of 8.8x and 4.5x over VITS-base, respectively.
The Mini FLY-TTS variant achieves a faster inference speed due to the fewer layers in the decoder.

The evaluation results on the audio quality are shown in Table \ref{table:mos_comparison}.
Our proposed FLY-TTS demonstrates comparable naturalness compared to the strong baseline MB-iSTFT, achieving a MOS score of 4.12 (versus 4.08) with fewer parameters and more efficient synthesis.
The minor decrease in MOS for Mini FLY-TTS is acceptable, given the fewer parameters.
The results of MCD show a similar tendency, with FLY-TTS having an MCD value (5.56) close to the VITS (5.49), while the Mini version is slightly higher.
Furthermore, our approach exhibits lower WERs than MB-iSTFT, indicating better intelligibility.
Overall, our proposed methods could maintain the audio quality while enjoying fewer parameters and faster inference.

\subsubsection{Ablation study}

Table \ref{table:ablation_study} details the ablation study.
When we replace the ConvNext-based decoder with MB-iSTFT \cite{mb-istft-vits}, the inference speed on both CPU and GPU decreases, mainly due to the upsampling units in the decoder of MB-iSTFT.
As detailed in Section \ref{sec:wavlm}, removing the WavLM discriminator from the adversarial network significantly reduces audio quality.
This, in turn, substantiates that FLY-TTS could leverage the acoustic and semantic information learned by self-supervised learning to facilitate generator updates, thereby enhancing the quality of synthesized speech.
The ablation study underscores the effectiveness of our proposed model.

\begin{table}[th]
    \caption{Ablation study of FLY-TTS.}
    \label{table:ablation_study}
    \centering
    \footnotesize
    \begin{tabular}{lcccc}
        \toprule
        \multirow{2}{*}{\textbf{Model}} & \multirow{2}{*}{\#\textbf{Params}} & \multirow{2}{*}{\textbf{MOS}(CI)} & \multicolumn{2}{c}{\textbf{RTF}}            \\
        \cmidrule(lr{0pt}){4-5}
                                        &                                    &                                   & CPU                              & GPU      \\
        \midrule
        Ground truth                    & -                                  & $4.21(\pm 0.10)$                  & -                                & -        \\
        \midrule
        FLY-TTS                         & $17.89$ M                          & $4.12(\pm 0.09)$                  & $0.0139$                         & $0.0062$ \\
        - ConvNeXt                      & $20.90$ M                          & $4.01(\pm 0.11)$                  & $0.0261$                         & $0.0098$ \\
        - WavLM                         & $17.89$ M                          & $3.98(\pm 0.10)$                  & $0.0136$                         & $0.0061$ \\
        \bottomrule
    \end{tabular}
\end{table}
\section{Conclusion}

In this paper, we present FLY-TTS, a new end-to-end text-to-speech (TTS) model designed to achieve fast and high-quality speech synthesis.
Built upon the VITS architecture, our model introduces several techniques to reduce model parameters and speed up inference,
such as grouped parameter-sharing and ConvNeXt-based decoder.
Additionally, we incorporate the pre-trained speech model WavLM as a discriminator to improve the quality of synthesized speech via adversarial training.
Experimental results on the LJSpeech dataset show that FLY-TTS can achieve comparable speech synthesis quality to strong baseline models,
while the model size and inference speed are significantly improved.
As part of future work, we consider expanding the model to multi-speaker scenarios and integrating style features to enable more diverse speech synthesis.

\clearpage
\tiny
\bibliographystyle{IEEEtran}
\bibliography{FLYTTS}

\begin{thebibliography}{10}
\providecommand{\url}[1]{#1}
\csname url@samestyle\endcsname
\providecommand{\newblock}{\relax}
\providecommand{\bibinfo}[2]{#2}
\providecommand{\BIBentrySTDinterwordspacing}{\spaceskip=0pt\relax}
\providecommand{\BIBentryALTinterwordstretchfactor}{4}
\providecommand{\BIBentryALTinterwordspacing}{\spaceskip=\fontdimen2\font plus
\BIBentryALTinterwordstretchfactor\fontdimen3\font minus \fontdimen4\font\relax}
\providecommand{\BIBforeignlanguage}[2]{{%
\expandafter\ifx\csname l@#1\endcsname\relax
\typeout{** WARNING: IEEEtran.bst: No hyphenation pattern has been}%
\typeout{** loaded for the language `#1'. Using the pattern for}%
\typeout{** the default language instead.}%
\else
\language=\csname l@#1\endcsname
\fi
#2}}
\providecommand{\BIBdecl}{\relax}
\BIBdecl

\bibitem{app9194050}
\BIBentryALTinterwordspacing
Y.~Ning, S.~He, Z.~Wu, C.~Xing, and L.-J. Zhang, ``A review of deep learning based speech synthesis,'' \emph{Applied Sciences}, vol.~9, no.~19, 2019. [Online]. Available: \url{https://www.mdpi.com/2076-3417/9/19/4050}
\BIBentrySTDinterwordspacing

\bibitem{tan2021survey}
X.~Tan, T.~Qin, F.~Soong, and T.-Y. Liu, ``A survey on neural speech synthesis,'' \emph{arXiv preprint arXiv:2106.15561}, 2021.

\bibitem{tacotron}
R.~Skerry-Ryan, E.~Battenberg, Y.~Xiao, Y.~Wang, D.~Stanton, J.~Shor, R.~Weiss, R.~Clark, and R.~A. Saurous, ``Towards end-to-end prosody transfer for expressive speech synthesis with tacotron,'' in \emph{international conference on machine learning}.\hskip 1em plus 0.5em minus 0.4em\relax PMLR, 2018, pp. 4693--4702.

\bibitem{glow-tts}
J.~Kim, S.~Kim, J.~Kong, and S.~Yoon, ``Glow-tts: A generative flow for text-to-speech via monotonic alignment search,'' \emph{Advances in Neural Information Processing Systems}, vol.~33, pp. 8067--8077, 2020.

\bibitem{vits}
J.~Kim, J.~Kong, and J.~Son, ``Conditional variational autoencoder with adversarial learning for end-to-end text-to-speech,'' in \emph{International Conference on Machine Learning}.\hskip 1em plus 0.5em minus 0.4em\relax PMLR, 2021, pp. 5530--5540.

\bibitem{naturalspeech}
X.~Tan, J.~Chen, H.~Liu, J.~Cong, C.~Zhang, Y.~Liu, X.~Wang, Y.~Leng, Y.~Yi, L.~He \emph{et~al.}, ``Naturalspeech: End-to-end text-to-speech synthesis with human-level quality,'' \emph{IEEE Transactions on Pattern Analysis and Machine Intelligence}, 2024.

\bibitem{mb-istft-vits}
M.~Kawamura, Y.~Shirahata, R.~Yamamoto, and K.~Tachibana, ``Lightweight and high-fidelity end-to-end text-to-speech with multi-band generation and inverse short-time fourier transform,'' in \emph{ICASSP 2023-2023 IEEE International Conference on Acoustics, Speech and Signal Processing (ICASSP)}.\hskip 1em plus 0.5em minus 0.4em\relax IEEE, 2023, pp. 1--5.

\bibitem{Wavlm}
S.~Chen, C.~Wang, Z.~Chen, Y.~Wu, S.~Liu, Z.~Chen, J.~Li, N.~Kanda, T.~Yoshioka, X.~Xiao \emph{et~al.}, ``Wavlm: Large-scale self-supervised pre-training for full stack speech processing,'' \emph{IEEE Journal of Selected Topics in Signal Processing}, vol.~16, no.~6, pp. 1505--1518, 2022.

\bibitem{ljspeech17}
K.~Ito and L.~Johnson, ``The lj speech dataset,'' \url{https://keithito.com/LJ-Speech-Dataset/}, 2017.

\bibitem{shen2018natural}
J.~Shen, R.~Pang, R.~J. Weiss, M.~Schuster, N.~Jaitly, Z.~Yang, Z.~Chen, Y.~Zhang, Y.~Wang, R.~Skerrv-Ryan \emph{et~al.}, ``Natural tts synthesis by conditioning wavenet on mel spectrogram predictions,'' in \emph{2018 IEEE international conference on acoustics, speech and signal processing (ICASSP)}.\hskip 1em plus 0.5em minus 0.4em\relax IEEE, 2018, pp. 4779--4783.

\bibitem{fastspeech}
Y.~Ren, Y.~Ruan, X.~Tan, T.~Qin, S.~Zhao, Z.~Zhao, and T.-Y. Liu, ``Fastspeech: Fast, robust and controllable text to speech,'' \emph{Advances in neural information processing systems}, vol.~32, 2019.

\bibitem{fastspeech2}
Y.~Ren, C.~Hu, X.~Tan, T.~Qin, S.~Zhao, Z.~Zhao, and T.-Y. Liu, ``Fastspeech 2: Fast and high-quality end-to-end text to speech,'' in \emph{International Conference on Learning Representations}, 2020.

\bibitem{NixTTS}
R.~Chevi, R.~E. Prasojo, A.~F. Aji, A.~Tjandra, and S.~Sakti, ``Nix-tts: Lightweight and end-to-end text-to-speech via module-wise distillation,'' in \emph{2022 IEEE Spoken Language Technology Workshop (SLT)}.\hskip 1em plus 0.5em minus 0.4em\relax IEEE, 2023, pp. 970--976.

\bibitem{Light-tts}
S.~Li, B.~Ouyang, L.~Li, and Q.~Hong, ``Light-tts: Lightweight multi-speaker multi-lingual text-to-speech,'' in \emph{ICASSP 2021-2021 IEEE International Conference on Acoustics, Speech and Signal Processing (ICASSP)}.\hskip 1em plus 0.5em minus 0.4em\relax IEEE, 2021, pp. 8383--8387.

\bibitem{SpeedySpeech}
J.~Vainer and O.~Dušek, ``{SpeedySpeech: Efficient Neural Speech Synthesis},'' in \emph{Proc. Interspeech 2020}, 2020, pp. 3575--3579.

\bibitem{Lightspeech}
R.~Luo, X.~Tan, R.~Wang, T.~Qin, J.~Li, S.~Zhao, E.~Chen, and T.-Y. Liu, ``Lightspeech: Lightweight and fast text to speech with neural architecture search,'' in \emph{ICASSP 2021-2021 IEEE International Conference on Acoustics, Speech and Signal Processing (ICASSP)}.\hskip 1em plus 0.5em minus 0.4em\relax IEEE, 2021, pp. 5699--5703.

\bibitem{EfficientSpeech}
R.~Atienza, ``Efficientspeech: An on-device text to speech model,'' in \emph{ICASSP 2023-2023 IEEE International Conference on Acoustics, Speech and Signal Processing (ICASSP)}.\hskip 1em plus 0.5em minus 0.4em\relax IEEE, 2023, pp. 1--5.

\bibitem{AdaVITS}
K.~Song, H.~Xue, X.~Wang, J.~Cong, Y.~Zhang, L.~Xie, B.~Yang, X.~Zhang, and D.~Su, ``Adavits: Tiny vits for low computing resource speaker adaptation,'' in \emph{2022 13th International Symposium on Chinese Spoken Language Processing (ISCSLP)}.\hskip 1em plus 0.5em minus 0.4em\relax IEEE, 2022, pp. 319--323.

\bibitem{Hifi-gan}
J.~Kong, J.~Kim, and J.~Bae, ``Hifi-gan: Generative adversarial networks for efficient and high fidelity speech synthesis,'' \emph{Advances in Neural Information Processing Systems}, vol.~33, pp. 17\,022--17\,033, 2020.

\bibitem{transformer}
A.~Vaswani, N.~Shazeer, N.~Parmar, J.~Uszkoreit, L.~Jones, A.~N. Gomez, {\L}.~Kaiser, and I.~Polosukhin, ``Attention is all you need,'' \emph{Advances in neural information processing systems}, vol.~30, 2017.

\bibitem{ALBERT}
Z.~Lan, M.~Chen, S.~Goodman, K.~Gimpel, P.~Sharma, and R.~Soricut, ``Albert: A lite bert for self-supervised learning of language representations,'' in \emph{International Conference on Learning Representations}, 2020.

\bibitem{universal}
M.~Dehghani, S.~Gouws, O.~Vinyals, J.~Uszkoreit, and L.~Kaiser, ``Universal transformers,'' in \emph{International Conference on Learning Representations}, 2019.

\bibitem{Portaspeech}
Y.~Ren, J.~Liu, and Z.~Zhao, ``Portaspeech: Portable and high-quality generative text-to-speech,'' \emph{Advances in Neural Information Processing Systems}, vol.~34, pp. 13\,963--13\,974, 2021.

\bibitem{Nanoflow}
S.-g. Lee, S.~Kim, and S.~Yoon, ``Nanoflow: Scalable normalizing flows with sublinear parameter complexity,'' \emph{Advances in Neural Information Processing Systems}, vol.~33, pp. 14\,058--14\,067, 2020.

\bibitem{WaveNet}
A.~{van den Oord}, S.~Dieleman, H.~Zen, K.~Simonyan, O.~Vinyals, A.~Graves, N.~Kalchbrenner, A.~Senior, and K.~Kavukcuoglu, ``{WaveNet: A Generative Model for Raw Audio},'' in \emph{Proc. 9th ISCA Workshop on Speech Synthesis Workshop (SSW 9)}, 2016, p. 125.

\bibitem{Vocos}
H.~Siuzdak, ``Vocos: Closing the gap between time-domain and fourier-based neural vocoders for high-quality audio synthesis,'' in \emph{The Twelfth International Conference on Learning Representations}, 2024.

\bibitem{convnext}
Z.~Liu, H.~Mao, C.-Y. Wu, C.~Feichtenhofer, T.~Darrell, and S.~Xie, ``A convnet for the 2020s,'' in \emph{Proceedings of the IEEE/CVF conference on computer vision and pattern recognition}, 2022, pp. 11\,976--11\,986.

\bibitem{valle}
C.~Wang, S.~Chen, Y.~Wu, Z.~Zhang, L.~Zhou, S.~Liu, Z.~Chen, Y.~Liu, H.~Wang, J.~Li \emph{et~al.}, ``Neural codec language models are zero-shot text to speech synthesizers,'' \emph{arXiv preprint arXiv:2301.02111}, 2023.

\bibitem{Audiolm}
Z.~Borsos, R.~Marinier, D.~Vincent, E.~Kharitonov, O.~Pietquin, M.~Sharifi, D.~Roblek, O.~Teboul, D.~Grangier, M.~Tagliasacchi \emph{et~al.}, ``Audiolm: a language modeling approach to audio generation,'' \emph{IEEE/ACM Transactions on Audio, Speech, and Language Processing}, 2023.

\bibitem{wav2vec2}
A.~Baevski, Y.~Zhou, A.~Mohamed, and M.~Auli, ``wav2vec 2.0: A framework for self-supervised learning of speech representations,'' \emph{Advances in neural information processing systems}, vol.~33, pp. 12\,449--12\,460, 2020.

\bibitem{mao2017least}
X.~Mao, Q.~Li, H.~Xie, R.~Y. Lau, Z.~Wang, and S.~Paul~Smolley, ``Least squares generative adversarial networks,'' in \emph{Proceedings of the IEEE international conference on computer vision}, 2017, pp. 2794--2802.

\bibitem{Styletts2}
Y.~A. Li, C.~Han, V.~Raghavan, G.~Mischler, and N.~Mesgarani, ``Styletts 2: Towards human-level text-to-speech through style diffusion and adversarial training with large speech language models,'' \emph{Advances in Neural Information Processing Systems}, vol.~36, 2024.

\bibitem{adamw}
I.~Loshchilov and F.~Hutter, ``Decoupled weight decay regularization,'' in \emph{International Conference on Learning Representations}, 2019.

\bibitem{Whisper}
A.~Radford, J.~W. Kim, T.~Xu, G.~Brockman, C.~McLeavey, and I.~Sutskever, ``Robust speech recognition via large-scale weak supervision,'' in \emph{International Conference on Machine Learning}.\hskip 1em plus 0.5em minus 0.4em\relax PMLR, 2023, pp. 28\,492--28\,518.

\end{thebibliography}

\end{document}